\documentstyle[aps,preprint]{revtex}
\input epsf
\title{Semiclassical stability of the extreme \\
  Reissner-Nordstr\"{o}m black hole}
\author{Paul R.  Anderson}
\address{Department of Physics\\Wake Forest University \\Winston-Salem, North
 Carolina 27109}
\author{William A.  Hiscock and Daniel J.  Loranz} \address{Department of
 Physics\\Montana State University\\Bozeman, Montana 59717, U.S.A.}

\begin{document}
\maketitle
\begin{abstract}

      The stress-energy tensor of a free quantized scalar field is
calculated in the extreme Reissner-Nordstr\"{o}m black hole spacetime in
the zero temperature vacuum state.  The stress-energy appears to be regular
on the event horizon, contrary to the suggestion provided by
two-dimensional calculations.  An analytic calculation on the event horizon
for a thermal state shows that if the temperature is nonzero then the
stress-energy diverges strongly there.

\end{abstract}
\pagebreak

     Extreme black holes have played an important role in recent
investigations of information loss in black hole evaporation, in studies of
black hole thermodynamics, and in superstring theory.  In the first case
models involving an extreme black hole have been studied \cite{hs} to avoid
the difficulties associated with quantum gravity at the endpoint of
evaporation.  If a zero temperature extreme black hole absorbs a small
amount of radiation its temperature will increase and its newly acquired
mass will radiate away via the Hawking effect, leaving a zero temperature
extreme black hole again in the final state.  This process may be
completely described within the context of a semiclassical theory, avoiding
the complications of Planck scale physics.  A black hole is traditionally
defined to be ``extreme'' if it has zero surface gravity; usually this is
taken to imply zero temperature.  However, in the context of black hole
thermodynamics, Hawking, Horowitz and Ross \cite{hhr} have recently
suggested that extreme black holes can be in thermal equilibrium with
radiation at an arbitrary temperature, and may have zero entropy despite
having nonzero horizon area.  Finally, extreme black holes may play a
crucial role in string theory, where they have been identified with massive
single string states \cite{s3,s4}.

      There is a potential problem with studying the properties of extreme
black holes because there are predictions that the stress-energy tensor for
quantized fields may diverge on the event horizons of such black holes.  In
two dimensions, Trivedi \cite{tr} has demonstrated that the trace anomaly
will cause a divergence of the stress-energy of a quantized field on the
horizon of a two-dimensional extreme black hole.  In four dimensions the
analytic approximations of Frolov and Zel'nikov \cite{fz} and Anderson,
Hiscock, and Samuel \cite{ahs} predict that the stress-energy tensors for
conformally invariant fields and for arbitrarily coupled massless scalar
fields respectively diverge on the event horizon of an extreme
Reissner-Nordstr\"{o}m black hole, for which the magnitude of the charge is
equal to the mass.  If the actual four dimensional stress-energy tensors
for these fields (rather than their two dimensional counterparts or their
analytical approximations) do diverge on the event horizons of extreme
black holes, this would imply that quantum effects substantially alter the
classical spacetime geometry around such black holes.

      Such divergences in the stress-energy tensor would imply that extreme
black holes are a third "venue" for quantum gravity (in addition to the
very early universe and the endpoint of black hole evaporation), despite
the fact that the spacetime curvatures near the event horizons of extreme
black holes with large masses are arbitrarily small.  While the precisely
extreme case may be physically inaccessible \cite{wi}, processes such as
black hole evaporation can yield states arbitrarily close to the extreme
state \cite{whl}.  If there is a physical divergence in the stress-energy
of a quantized field for the extreme case, then presumably there must be an
arbitrarily large stress-energy on the horizon of a nearly extreme black
hole ~\cite{r10}.

      In this letter we present an analytic calculation which shows that
the stress-energy tensor of a free conformally invariant scalar field
diverges on the event horizon of an extreme Reissner-Nordstr\"{o}m black
hole if the field is in a thermal state at any nonzero temperature.  We
also present the results of a numerical computation of the stress-energy
tensor for a free quantized massless scalar field with arbitrary curvature
coupling in the extreme Reissner-Nordstr\"{o}m spacetime with the field in
the zero-temperature Euclidean vacuum state.  These results provide strong
evidence that the stress-energy is finite on the event horizon if the field
is in the zero-temperature Euclidean vacuum state.  Thus we find for the
case of free quantized scalar fields that extreme Reissner-Nordstr\"{o}m
black holes can exist if the fields are in a zero temperature vacuum state,
but that they cannot exist in thermal equilibrium with radiation at a
nonzero temperature.

      The extreme Reissner-Nordstr\"{o}m spacetime in four dimensions may
be described by the metric \cite{units}
\begin{equation}
d s^2 = - {\left(1- {M \over r} \right)}^2 d t^2 + {\left(1 - {M \over r}
	\right)}^{-2} d r^2 + r^2 d \Omega^2 \;\;\; ,
\end{equation}
where the horizon is located at
$r=M$.  The coordinate system of Eq.(1) is singular at the horizon; in
order to determine whether a stress-energy tensor is regular on the horizon
it is necessary to set up a frame which is regular there.  Evaluating the
stress-energy tensor components in an orthonormal frame associated with a
freely falling observer, one finds that the stress-energy will be regular
at the horizon if and only if the components ${T_t}^t, {T_r}^r $, and
${T_\theta}^\theta$ (note that ${T_\phi}^\phi = {T_\theta}^\theta$), and
the quantity
\begin{equation}
F = \left( {T_r}^r - {T_t}^t \right) {\left(1
	- {M \over r} \right)}^{-2}
\end{equation}
are regular as $r \rightarrow M$.  Both the energy density and radial
pressure as measured by an infalling observer will be proportional to
$F$ near the horizon.

      The previously predicted divergences occur for the quantity $F$ in
Eq.(2) when the fields are in the zero temperature vacuum state.  For a
conformally invariant scalar field in a two dimensional extreme
Reissner-Nordstr\"{o}m spacetime \cite{tr,whd}
\begin{equation}
F = - {M\over {6 \pi r^2 (r - M) }} \;\;\; .
\end{equation}
For an arbitrarily coupled massless scalar field in a four dimensional
extreme Reissner-Nordstr\"{o}m spacetime the analytic approximations
for the stress-energy tensor give \cite{fz,ahs}
\begin{equation}
F = {{14 M^3 - 11 M^2 r} \over {360 \pi^2 r^6 (r-M)}} -
	{{M^2} \over {60 \pi^2 r^6}}\,\, ln \left[{\mu
	\over {2 r}} (r-M) \right ] \;\;\; ,
\end{equation}
where $\mu$ is an arbitrary positive constant.

      At the event horizon of the extreme Reissner-Nordstr\"{o}m black hole
the ($t,r,\theta,\phi$) components of the stress-energy tensor for a
conformally invariant field can be computed analytically.  This is because
the geometry at the horizon is asymptotically congruent to that of the
conformally flat Bertotti-Robinson electromagnetic spacetime
\cite{r13,r14,r15}.  The Bertotti-Robinson metric may be written as
\begin{equation}
ds^2 = M^2 {\bar r}^{-2} (-d t^2 + d {\bar r}^2 + {\bar
	r}^2 d \Omega^2 ) \;\;\; .
\end{equation}
The manner in which the extreme Reissner-Nordstr\"{o}m metric approaches
that of Eq.(5) as $r \rightarrow M$ may be seen by expanding the metric
coefficients in Eq.(1) in a power series about $r = M$ and then setting
$\bar r = M^2/(r-M)$.  Brown and Cassidy \cite{bc} and Bunch \cite{bu}
have derived the following expression for the stress-energy tensor of
a conformally invariant quantized field in a conformally flat spacetime:
\begin{eqnarray}
{\langle T_{\mu\nu}\rangle}_{ren} & = & (\tilde g /g)^{1/2}
	{\langle T_{\mu \nu}[\tilde g_{\kappa \lambda}]\rangle}_{ren}
	 + {\alpha \over 3} \left( g_{\mu \nu} {R^{;\kappa}}_{;\kappa}
	 - R_{;\mu \nu} + R R_{\mu \nu} - {1 \over 4} g_{\mu
	\nu} R^2 \right) \nonumber \\ & & + \beta \left({2 \over 3}
	R R_{\mu \nu} -{R_\mu}^\kappa R_{\kappa \nu} + {1 \over 2}
	g_{\mu \nu} R_{\kappa \lambda}R^{\kappa \lambda} - {1 \over 4}
	g_{\mu \nu} R^2 \right)
\end{eqnarray}
Here $\tilde g$ is the flat space metric and $\langle T _{\mu \nu} [\tilde
g] \rangle $ is the stress-energy tensor of the quantized field in the flat
space; for the conformally invariant scalar field $\alpha = \beta = (2880
\pi^2)^{-1}$.  Combining Eqs.(5,6) with a thermal state in Minkowski space,
we find the stress-energy tensor of a conformally invariant scalar field in
the Bertotti-Robinson spacetime at temperature $T$ is \cite{r18}
\begin{equation}
	{\langle {T _ \mu}^\nu \rangle}_{ren} = {1 \over {2880
	\pi^2 M^4}} diag (1,1,1,1) + {M^4 \over {(r-M)^4}}
	{{\pi^2 T^4} \over 30}diag (-1, {1 \over 3}, {1 \over 3},
	{1 \over 3} )
\end{equation}
This is clearly divergent in the limit $r \rightarrow M $ if $T \neq 0$.

      If $T = 0$, then the stress-energy tensor in the Bertotti-Robinson
spacetime is regular everywhere.  However, this does not guarantee that the
stress-energy is finite on the event horizon in the extreme
Reissner-Nordstr\"{o}m spacetime.  This is because the Reissner-
Nordstr\"{o}m metric only asymptotically approaches the Bertotti-Robinson
one; the quantity $F$ in Eq.(2) may be nonzero on the event horizon of the
extreme black hole, although it is precisely zero in the Bertotti-Robinson
spacetime.

      To investigate whether there is a divergence on the event horizon in
the T = 0 case we have applied the method developed by Anderson, Hiscock
and Samuel \cite{ahs} for calculating the stress-energy tensor of quantized
scalar fields in static spherical spacetimes to the case of a massless
arbitrarily coupled field in the extreme Reissner- Nordstr\"{o}m spacetime.
Using these techniques, one may compute the stress-energy tensor components
at any value of r outside the horizon \cite{r19}.

      Symbolically, the stress-energy tensor components may be divided into
conformal and nonconformal contributions in terms of two tensors,
${C_\mu}^\nu $ and ${D_\mu}^\nu$:  where $\xi$ is the curvature coupling
constant for the scalar field.  The values of ${C_t}^t, {C_r}^r$, and
${C_\theta}^\theta$ extrapolated to the horizon are the same, within the
accuracy of our computations (approximately 5 digits near the horizon for
these components), as those obtained from the analytic calculation for the
Bertotti-Robinson spacetime, Eq.(7).  The behavior of the components of
${C_\mu}^\nu$ is illustrated in Fig.  (1).  The values of the components
are clearly finite as $r \rightarrow M$ and they smoothly approach the
Bertotti-Robinson values on the horizon.  The behavior of the corresponding
components of ${D_\mu}^\nu$ is illustrated in Fig.  (2).  They approach
zero as $r \rightarrow M$ and are thus finite also.

      In order to determine the regularity of the stress-energy tensor on
the horizon, we must also examine the quantity $F$ defined in Eq.(2).  This
quantity is plotted in Figures (3) and (4) for the contributions of
${C_\mu}^\nu $ and ${D_\mu}^\nu$ respectively.  In Figure (3) we also show
$F$ for the conformally invariant scalar field for the two dimensional
Reissner-Nordstr\"{o}m black hole and the four dimensional analytic
approximation with $\mu=2$.  Our data points are explicitly displayed for
the numerically computed curves to give the reader a better idea of the
quality of the information; we have, however, suppressed the dots
representing our calculated points at $r = 1.002M, 1.003M$, and $1.004M$
for clarity.  The numerically computed curves stop at our innermost data
point, $r = 1.001 M$, since we cannot directly compute the value of $F$ on
the horizon.

Inspection of these figures clearly shows that no divergence at the horizon
is to be expected.  This may be made quantitative by fitting our data
points to a series in the dimensionless quantity $s = (r-M)/M$ and
comparing the magnitude of the coefficient of hypothetical $s^{-1}$ and
$ln(s)$ terms with other finite terms.  Fits may be tried with different
sets of points and different numbers of terms in the series to obtain a
robust estimate of the power in an $s^{-1}$ term.  Such calculations all
yield a coefficient for the $s^{-1}$ term which is at most of order
$10^{-4}$ times the coefficient of the constant term.  In contrast from
Eq.(3) the ratio of these coefficients for the two- dimensional
stress-energy tensor is $1/2$, while from Eq.(4) for the analytical
approximation (with $\mu=2$ chosen for simplicity) the ratio is $3/29$.  As
an example of a typical fit, if our four innermost data points, at $s =
0.001, 0.002, 0.003$, and $0.004$, are fitted to an expansion in powers of
$s$, including divergent $s^{-1}$ and $ln(s)$ terms, we obtain for the
tensor ${C_\mu}^\nu$ the fit
\begin{equation}
	90 (8 M)^4 \pi^2 \left({C_r}^r - {C_t}^t \right)
	{\left(1 - {M \over r} \right)}^{-2} = {0.8122
	\over s} + 1333 ln(s) + 33600 - 435000 s \;\;\; .
\end{equation}
Since our numerical calculations of $F$ near the event horizon are accurate to
between two and three digits, it is clear that all the power in the
$s^{-1}$ term is simple "numerical noise", not to be taken seriously.

      Given that no divergence of the form $s^{-1}$ exists, one can try to
fit a series with just a $ln(s)$ term plus terms which are finite on the
horizon.  In these cases we find the coefficient of the $ln$ term is
generally at most $10^{-2}$ times the coefficient of the constant term.  In
contrast, from Eq.(4) this ratio is $6/29$ in the analytic approximation.

      Thus, numerical calculations strongly suggest that the divergence in
the two- dimensional black hole stress-energy tensor discovered by Trivedi
and those in the four dimensional analytic approximations to the
stress-energy tensor do not exist in the actual four dimensional
stress-energy tensor, at least in the case of an extreme Reissner-
Nordstr\"{o}m spacetime.  Thus real, physical extreme black holes will not
have divergent vacuum stress-energy on their horizon if they are in the
zero temperature vacuum state.  However, our analytic calculations also
show that the stress-energy of free quantized fields does diverge on the
event horizon of an extreme Reissner-Nordstr\"{o}m black hole if the fields
are in a nonzero temperature thermal state.  Thus, in this case, the
suggestion of Hawking, Horowitz, and Ross \cite{hhr} that extreme black
holes can be in thermal equilibrium with radiation at an arbitrary
temperature is incorrect.

\acknowledgements

      P.R.A.  would like to thank J.  S.  Dowker for a helpful
conversation.  The work of W.A.H.  was supported in part by National
Science Foundation Grant No.  PHY92-07903.

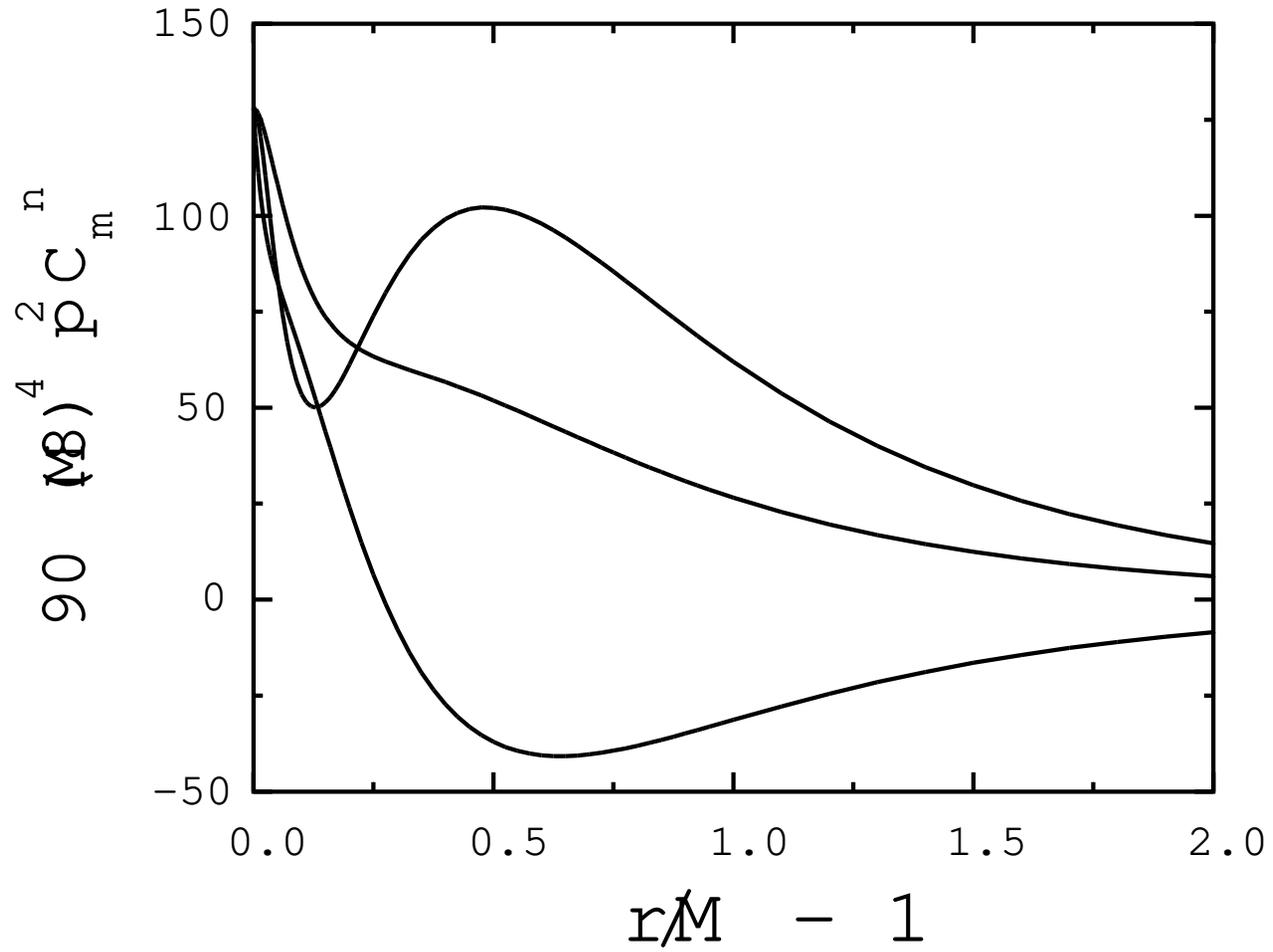
\begin{figure}
\caption{ The curves in this figure display the values of
${C_t}^t$, ${C_r}^r$, and ${C_\theta}^\theta$ (from top to bottom at $r =
2M$), for massless scalar fields around an extreme Reissner-Nordstr\"{o}m
black hole.}
\end{figure}

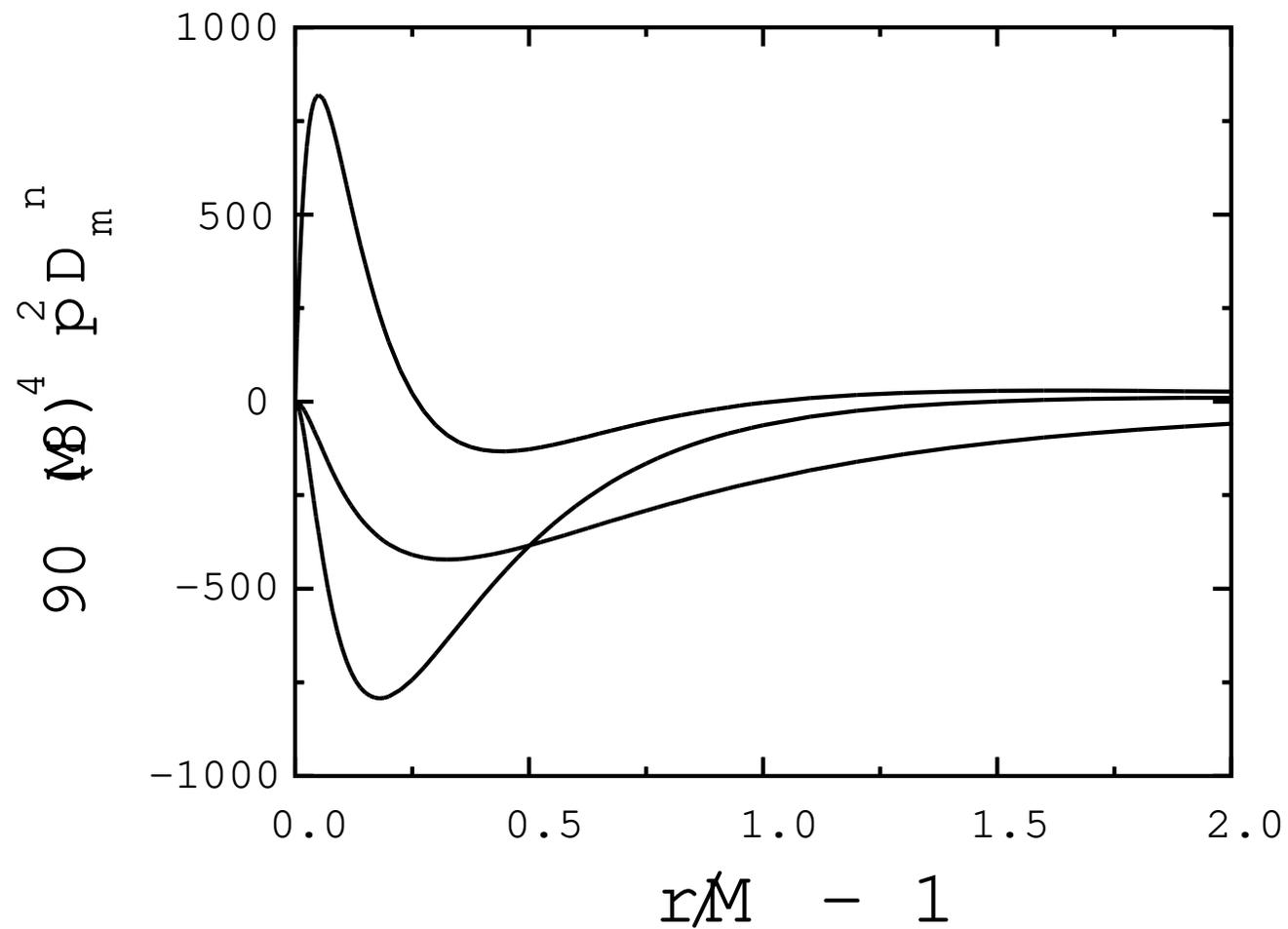
\begin{figure}
\caption{ The curves in this figure display the values of
${D_\theta}^\theta$, ${D_t}^t$, and ${D_r}^r$ (from top to bottom at $r =
2M$), for massless scalar fields around an extreme Reissner-Nordstr\"{o}m
black hole.}
\end{figure}

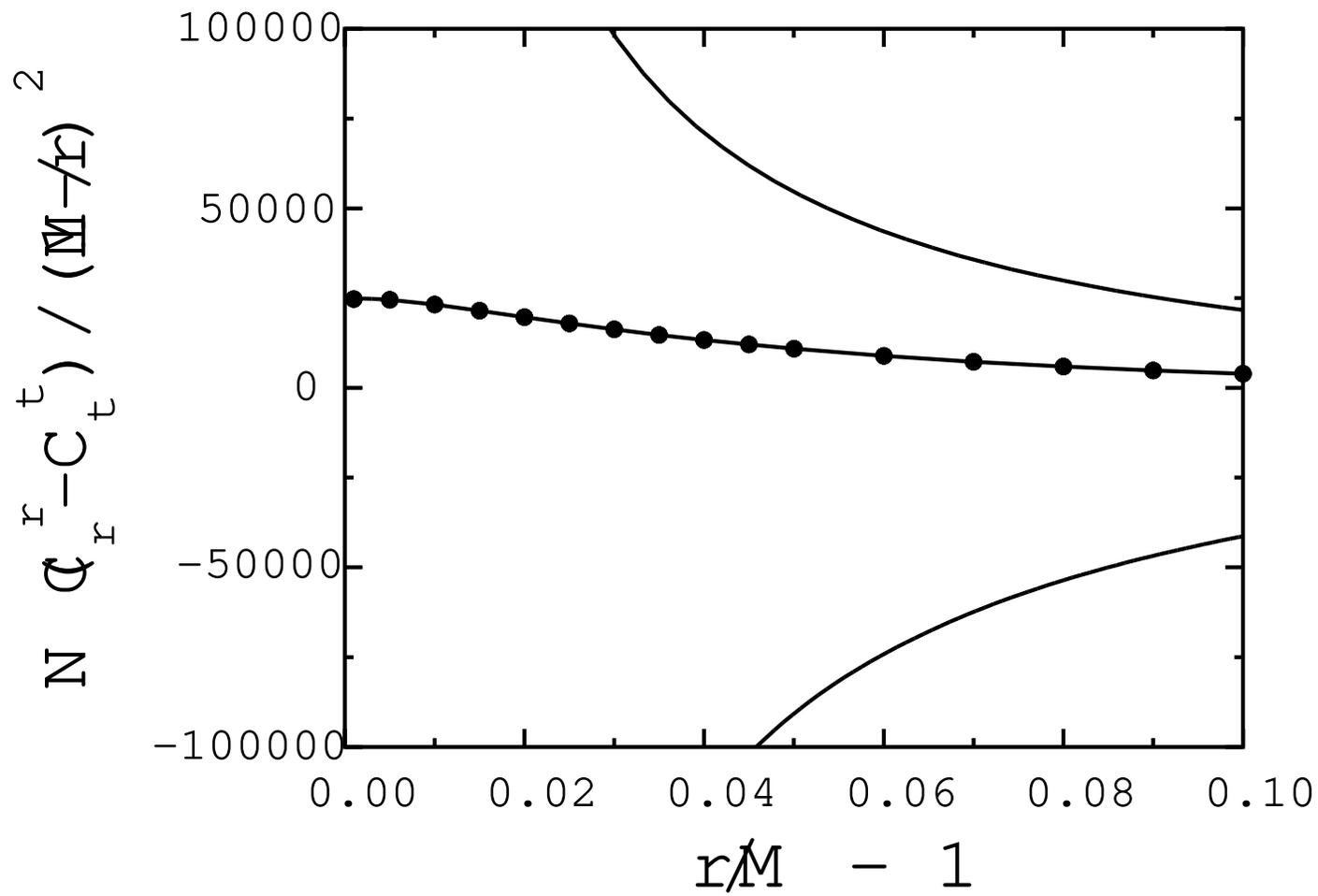
\begin{figure}
\caption{ These curves display the value of the quantity
$N({C_r}^r - {C_t}^t)/(1 - M/r)^2$ for a massless scalar field around an
extreme Reissner-Nordstr\"{o}m black hole.  From top to bottom, the curves
represent the four dimensional analytic approximation, the numerically
computed four dimensional values and the two-dimensional black hole result.
The constant $N$ has the value $90(8M)^4 \pi^2$ for the four dimensional
curves and $30000 \pi M^2$ for the two dimensional curve.  No divergence at
the horizon $(r=M)$ is apparent for the numerically computed curve.}
\end{figure}

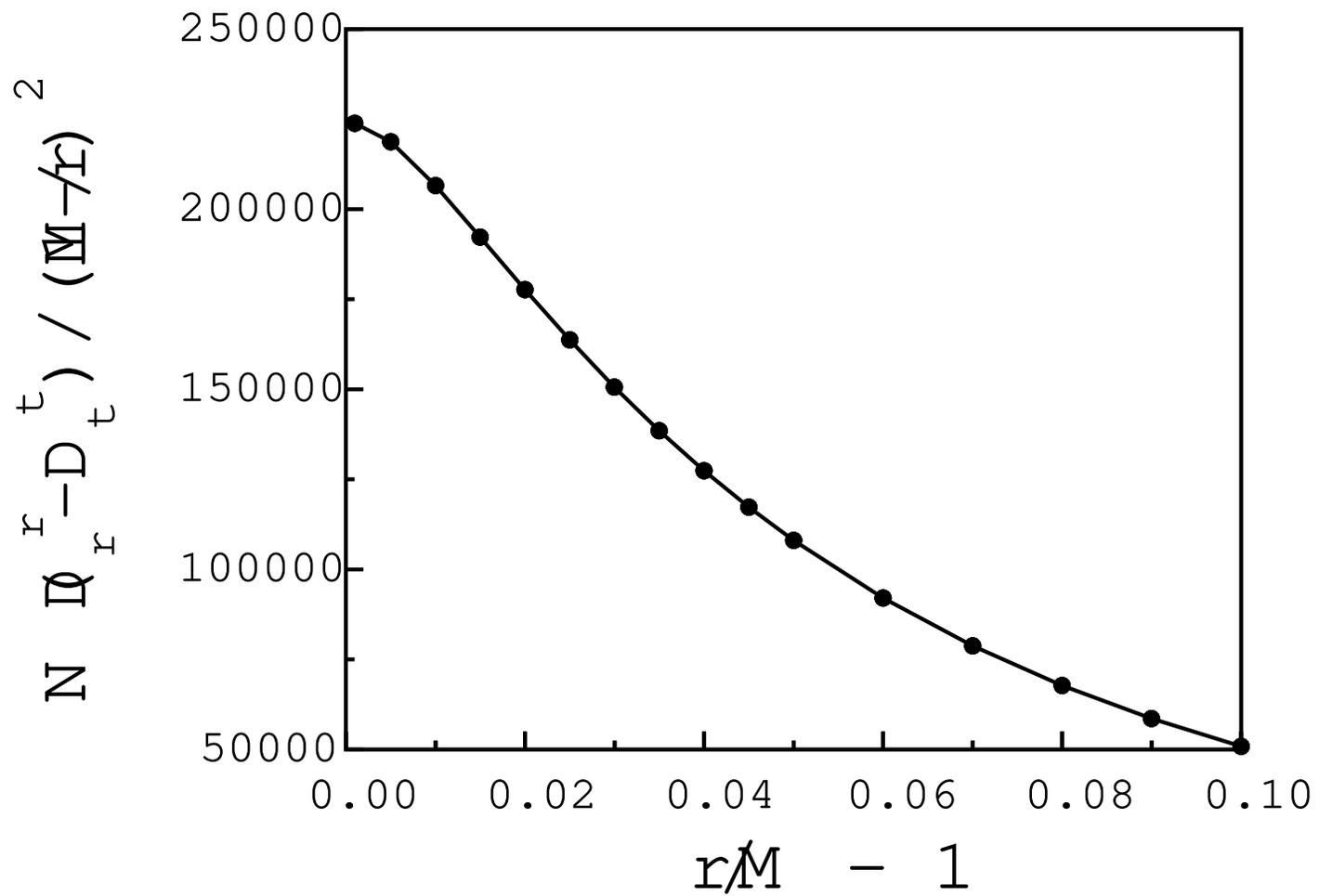
\begin{figure}
\caption{ The curve displays the value of the quantity
$N({D_r}^r - {D_t}^t)/(1-M/r)^2$ for a massless scalar field around an
extreme Reissner-Nordstr\"{o}m black hole.  The constant $N$ has the value
$90(8M)^4 \pi^2$.  No divergence at the horizon $(r=M)$ is apparent.}
\end{figure}


\begin{references}
\bibitem{r10} This is
      certainly true for two dimensional black holes; see Ref.[12].
\bibitem{r19} The components contain some terms (all finite on the
      horizon) which depend on an arbitrary renormalization parameter; we
      have used the same numerical value for this parameter as in Ref.  [7]
      where this issue is discussed in detail.
\end{references}
\end{document}